\documentclass{sigchi-ext}
\usepackage[T1]{fontenc}
\usepackage{textcomp}
\usepackage[scaled=.92]{helvet} 
\usepackage{graphicx} 
\usepackage{balance}  
\usepackage{booktabs} 
\usepackage{ccicons}  
\usepackage{ragged2e} 

\def\plaintitle{Opportunity in Conflict: Understanding Tension Among Key Groups on the Trail} 
\def\emptyauthor{}
\def\plainkeywords{Hiking, Technology, Trail}

\title{Opportunity in Conflict: Understanding Tension Among Key Groups on the Trail}

\numberofauthors{4}
\author{%
  \alignauthor{%
    \textbf{Lindah Kotut}\\
    \affaddr{Department of Computer Science}\\    
    \affaddr{Virginia Tech} \\
    \affaddr{Blacksburg, VA 24061, USA} \\
    \email{lkotut@vt.edu} }\alignauthor{%
    \textbf{D. Scott McCrickard}\\
    \affaddr{Department of Computer Science}\\   
    \affaddr{Virginia Tech} \\
    \affaddr{Blacksburg, VA 24061, USA} \\
    \email{mccricks@cs.vt.edu} } \vfil \alignauthor{%
    \textbf{Michael Horning}\\
    \affaddr{Department of Communication}\\    
    \affaddr{Virginia Tech} \\
    \affaddr{Blacksburg, VA 24061, USA} \\
    \email{mhorning@vt.edu} } \vfil \alignauthor{%
    \textbf{Steve Harrison}\\
    \affaddr{Department of Computer Science}\\ 
    \affaddr{Virginia Tech} \\
    \affaddr{Blacksburg, VA 24061, USA} \\
    \email{sharrison@vt.edu} } }

\definecolor{linkColor}{RGB}{6,125,233}
\hypersetup{%
  pdftitle={\plaintitle},
  pdfauthor={\emptyauthor},
  pdfkeywords={\plainkeywords},
  bookmarksnumbered,
  pdfstartview={FitH},
  colorlinks,
  citecolor=black,
  filecolor=black,
  linkcolor=black,
  urlcolor=linkColor,
  breaklinks=true,
}

\begin{document}

\CopyrightYear{2018}
\doi{http://dx.doi.org/10.1145/2858036.2858119}

\maketitle

\RaggedRight{} 


\begin{abstract}
  UPDATED---\today. This paper examines the question of who technology users on the trail are, what their technological uses and needs are, and what conflicts exist between different trail users regarding technology use and experience, toward understanding how experiences of trail users contribute to designers. We argue that exploring these tensions provide opportunities for design that can be used to both mitigate conflicts and improve community on the trail.
\end{abstract}



\section{Introduction}
People tend to make assumptions about the trail: who uses trails, what technology they use, and their attitudes toward the usage of said technologies \cite{Bryson1998}. Our approach contends that understanding trail users and their dynamics, particularly the tensions between different hiker groups, helps with understanding how these groups use technology. This understanding will help in directing analysis and presenting design guidance and/or opportunities for encouraging community, toward diffusing inter-group conflict. 

Tension can exist in the roles of groups on the trail. Hunters, for example, agree on the ethos of ``fair chase" \cite{Su2017}, but different subgroups differ on how they interpret this notion depending on their attitude towards the role of weapon technology (crossbows vs bows, rifles vs muzzleloaders, rifles vs bows) in hunting. The role of technology enhances personal experience on the trail, such as the use of fitbits and headphones \cite{Anderso2017}; citizen scientist water quality monitoring \cite{Rapousis2015}, and logistical planning of trail practicalities (e.g., campsite reservations, rest-room facilities).

\marginpar{\vspace{-12pc}
\textbf{TEAMS}\\
The participants were divided into two teams: \textbf{ \color{blue} Blue Team} used blue sticky-notes. The  \textbf{\color{yellow} Yellow Team} used yellow sticky-notes. \\ 

\textbf{\\ VOTING}\\
\textbf{\color{green}Green dots:} were used by participants to vote for hiker-groups considered to benefit from technology. \\
\textbf{\color{red}Red dots:} were used to vote for groups that do not benefit from technology \\
}

\begin{marginfigure}[5pc]
  \begin{minipage}{\marginparwidth}
    \centering
    \includegraphics[width=0.6\marginparwidth]{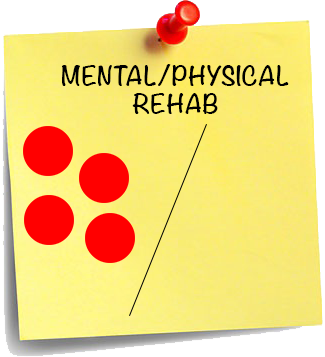}
    \caption{\textit{Yellow Team} made a distinction between \textit{mental and physical rehab} group in contrast to viewing it one cohesive group.}
      ~\label{fig:yellow-mental}
  \end{minipage}
\end{marginfigure}

\section{Approach}
Our approach builds upon three previous affinity diagramming sessions used to identify different facets of roles and goals for technology on the trail. The first session involved 25 participants who were tasked in identifying \textit{who} the trail users are. From this session, 132 different user roles were identified. The second affinity diagramming session involving 9 participants to determine \textit{why} these groups are on the trail based on common goals identified by clustering these groups. 

A third session engaged 10 participants divided into 2 teams (coded blue and yellow team) of 5 people, each focusing on 35 unique hiker groups from the original session (table \ref{tab:table1}), approached the question of \textit{what} technology design opportunities could be found in these groups and goals, by considering the question of \textit{benefit}: which groups benefit from technology, and which do not? Each participant in the team was given 8 votes: 4 (indicated with green dots) used to signify groups they judged to benefit from technology, the remaining 4 (red dots) to signify a detriment. 

This third session enabled a prioritization of user needs based not only on those groups identified to most benefit from technology, but also on groups that revealed tensions and conflict. This paper describes interesting groups that emerged from the exercise and further, how different tensions surfaced. Finally, we consider design opportunities proffered by these tensions. 

\begin{figure*}
  \centering
  \includegraphics[width=2.0\columnwidth]{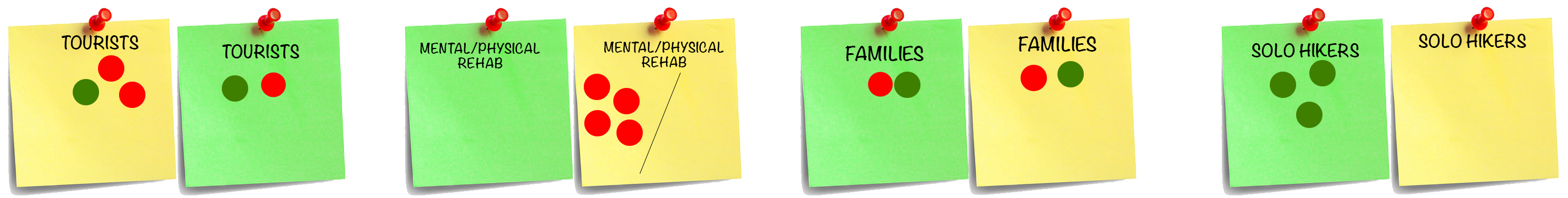}
  \caption{Selection of trail users considered contentious either explicitly based on vote discrepancy or implicitly based on team discussions.}~\label{fig:vote-grid}
\end{figure*}

\begin{table}
  \centering
  \begin{tabular}{l r r}
    \toprule
      Activists & Historians & Property Owners  \\
      Bikers/Activists & Hunters & Recreational \\
      Bird Watchers & Locals & Retirees \\
      Boy/Girl Scouts & Loggers & \textbf{Scientists} \\
      Day Hikers & Maintenance Workers & \textbf{S and R Workers} \\
      Exercisers & \textbf{Mental/Physical Rehab} & Section Hikers  \\
      \textbf{Families} & Horse-Back Riders & \textbf{Solo Hikers} \\
      \textbf{Farmers} & Park Rangers & Sponsored Hikers \\
      Firemen & Pet Owners & \textbf{Tourists} \\
      Fishermen & Picnickers  & Thru Hikers \\
      \textbf{Guide-Book Authors} & Plant Foragers & Trail Angels \\
      Herbalists & Prof/ Army Training & 	\\
    \bottomrule
  \end{tabular}
  \caption{35 unique groups were curated from previously identified hiker roles and used to determine technological benefits for each. }~\label{tab:table1}
\end{table}

\section{Observations}

Figure \ref{fig:vote-grid} showcases how each team voted across the groups for the most contentious user groups. Two groups particularly stood out from this exercise, based on the vote discrepancy across teams: \textit{Mental and Physical Rehab}, which received 4 red votes from the yellow team (though no votes from the blue team), and \textit{Solo Hikers} which received 3 green votes from the blue team (though no votes from the yellow team). 

After the clustering and voting exercise, each team was tasked with selecting a group of hikers considered to be interesting or contentious based on voting decisions and group debate. These groups were then discussed amongst all the ten participants to understand inter-group tensions and possible design opportunities. 

\begin{marginfigure}[0pc]
  \begin{minipage}{\marginparwidth}
    \centering
    \includegraphics[width=0.9\marginparwidth]{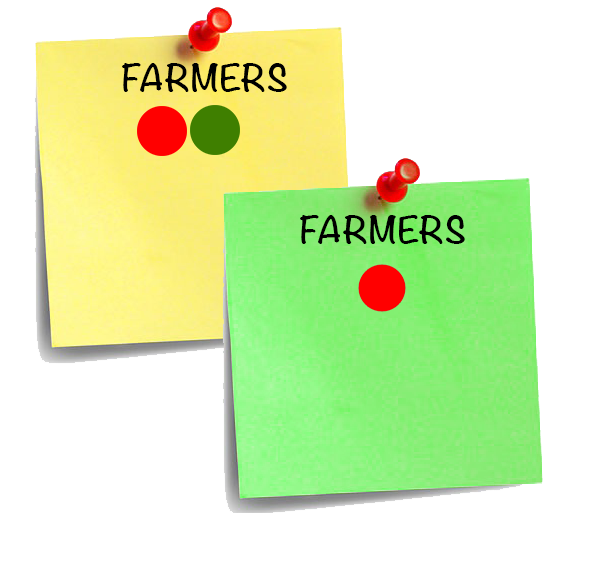}
    \caption{How the two teams voted for the case of \textit{Farmers} benefiting from technology. Both teams selected this group as contentious.}
      ~\label{fig:farmervotes}
  \end{minipage}
\end{marginfigure}

\section{Interesting Groups}
A selection of hiker groups emerged as interesting-- from how they were organized in clusters, how the teams voted for them, and how groups were selected by each group for further discussions. The selected groups were connected with interesting characteristics and underlying issues that led to the choices.

\subsection{(Un)Clear Hiking Goals}
The \textit{Families} group was selected by both teams as interesting not only in how to design for them, but also in how the group sparked debate on the difference between interacting with technology in contrast to with people on the trail, especially when the hiking goal is not clear. This discussion was also true of \textit{Tourists} group, which received the most votes across the two teams and further prompted a debate on the definition of `tourist'.  For these groups, often there are conflicting goals within the families or tourists, and often the goals are more ephemeral and not tied to reaching a destination, collecting artifacts, or completing a task.

\subsection{(Im)Practicality}
When discussing usefulness and practicality of technology on the trail, \textit{Search \& Rescue} group was voted for the most group likely to benefit -- notably because of direct association of the service with technology. This conclusion was also realized with the \textit{Scientists} group, tied with \textit{Search \& Rescue} in the number of votes received. All of these groups tend to have clearly defined goals that they wish to accomplish on the trail.

\subsection{Assisting vs Inhibiting}
An unanticipated though interesting discussion emerged when considering the \textit{Physical/Mental Rehab} group: the yellow team made a distinction between the mental and the physical elements of the group, thus initiating a question regarding the efficacy of current technology and possible technological innovations and applications to be used for purposes of mental rehab on the trail. Further, a debate on whether technology would benefit or inhibit the experience of this group on the trail. Technology can be tied to mental stresses, suggesting that it should be avoided on trails.

\section{Identifying Tensions and Design Opportunities}
The exercise and the discussions revealed common use patterns that present opportunities for design, and also common themes that reveal tensions between and within groups.

\subsection{Presence vs. Distraction}
The most explicit source of tension/conflict between groups are technology that distract from the moment: Email, social media, notifications, etc. that are considered to negatively impact \textit{Tourists}, contrasted with those that undermine the trail experience of a hiker-group altogether like \textit{Mental/Physical Rehab}. The yellow team were specific in differentiating between Mental and Physical aspects of rehab (Figure \ref{fig:yellow-mental}), and vociferous in their opposition to technology because of the negative effects on mental well-being.

\subsection{Experiential vs. Practical}
Tension emerged between groups where the line between experiential and practical gains was blurred: \textit{Families} is one group where the debate was whether the benefit the family gains from spending time chronicling the trail experience detracted from the experience of spending time with each other. The debate on \textit{Guide-Book Authors} considered the redundancy of \textit{Guide-books} with the popularity of online guides, against a preference for technology-agnostic alternatives for some users on the trail.

\subsection{Professional vs Amateur}
User expertise level mattered in the discussions about whether they would benefit from technology or not. This was reflected in the votes for groups that would benefit from technology: \textit{Search and Rescue Workers, Scientists and Hunters}, as there was a perceived distinction on the expertise of these users and in how technology assisted in acquitting their work.

\subsection{Known vs. The Unknown}
We acknowledge that the cause of tension between understood groups and those not well known. Our affinity diagramming sessions reflect the areas that are well known by the participants, particularly topics of interest to multiple people. This phenomenon was especially evident in the contrast between groups that got all green votes compared to those groups that did not receive any votes (e.g., \textit{Solo Hikers}, or groups that received one vote from a knowledgeable participant that did not inspire others to vote for it (e.g., \textit{Hunters}). We also note the explicit cases where the teams self-identify groups of hikers of which they do not fully grasp the breadth of what is involved in the technology: The \textit{Farmers} fell under this latter case.

\section{Future Plans}
Our pursuit of approaches for designing for the trail necessitate understanding the tension between trail users and how they interact with technology. Based on the discussions, we underscore the importance of distinguishing between experiencing the trail and assisting on the trail. Planning with this consideration in mind allows us to design to assist and augment the enterprise, without detracting from the experience.  

Future exercises should consider explicit tensions that may follow the common themes we have discussed in this paper that would further serve to inform further design opportunities. We feel that the tensions that were identified should lead to focus groups with key stakeholders, rich persona identification that highlights a depth of features, and scenarios of use that provide narrative descriptions of technology on the trail.


\balance{} 

\bibliographystyle{SIGCHI-Reference-Format}
\bibliography{extended-abstract}

\end{document}